\documentstyle[aps,epsfig]{revtex}
\begin{document}
\title{Note on chaos and diffusion}
\author{C. P. Dettmann}
\address{University of Bristol, Department of Mathematics, Bristol BS8 1TW, UK}
\author{E. G. D. Cohen}
\address{The Rockefeller University, 1230 York Ave, New York NY 10021, USA}
\date{\today}
\maketitle

\begin{abstract}
Using standard definitions of chaos (as positive Kolmogorov-Sinai entropy)
and diffusion (that multiple time distribution functions are Gaussian),
we show numerically that both chaotic and nonchaotic systems exhibit
diffusion, and hence that there is no direct logical connection between the
two properties.  This extends a previous result for two time distribution
functions. 
\end{abstract}

\section{Introduction}
It is generally surmised that microscopic chaos is a necessary condition
for diffusive behavior of a system.  Microscopic chaos here means a
positive Kolmogorov-Sinai (KS) entropy, or equivalently, at least one
positive Lyapunov exponent.  While this seemed to be confirmed by an
ingenious experiment with a Brownian particle~\cite{Gaspard}, we showed
that the same behavior on which the microscopic chaoticity of the Brownian
system was established also held for a manifestly nonchaotic system with
zero KS entropy~\cite{DCvB,DC}, so that both systems exhibited diffusion.%
\footnote{There are at least two integrable models (hence, with zero
Lyapunov exponents) that have been shown to exhibit diffusion as defined
by Eq.~\ref{e:multi}, a one dimensional hard rod model~\protect\cite{LPS}
and a heavy particle in a lattice of light particles~\protect\cite{R}.
In both cases there are an infinite number of moving particles, so we expect
that the KS entropy is infinite, as in the infinite ideal
gas~\protect\cite{GLA}.
Here we have only one moving particle, and we argued in Ref.~\cite{DC} that
the KS entropy of our nonchaotic models is zero.} 
In these works we used the mean square displacement (Eq.~\ref{e:msd} below), a
two time distribution function, as our definition of diffusion.
 	  
An anonymous referee of a Ref.~\cite{DC}, alerted us to a definition of
diffusion (Eq.~\ref{e:multi}) which involved multitime distribution functions
and the question arose whether those could distinguish between chaotic and
nonchaotic models, in that all our chaotic models and none of our nonchaotic
models would satisfy this definition.  In this note we show that this
behavior does not occur: no difference has been found between the diffusive
behavior of microscopically chaotic and nonchaotic models even in the 
multitime distribution functions we studied.

Our models, a subset of those of Ref.~\cite{DC},
consist of a point particle moving with unit speed in two
dimensions, undergoing specular reflections (collisions) with fixed
non-overlapping
scatterers.  The shapes, positions, and orientations of the scatterers
in the various models are defined in the next section, and can be classified
as Lorentz or (modified) Ehrenfest models.  For each fixed configuration
of scatterers, the particle's position ${\bf x}$ is determined numerically
as a function of the time $t$ for random initial positions ${\bf x}(0)$
(which are uniformly distributed over the plane, except that they cannot
lie inside a scatterer) and random initial velocity directions (also
uniformly distributed) of the particle.  The probabilities below are
defined with respect to this distribution of initial conditions.

The definition of diffusion proposed by the referee and used in this paper
is stated in terms of the probability of the particle initially at the origin
visiting $n$ regions at $n$ different specified times.  Explicitly, a system
is said to be diffusive, if, for all $n=1,2\ldots$, the $n+1$ time
(including $t=0$) probability distributions
${\cal P}$ satisfy the relation~\cite{referee}:
\begin{eqnarray}
\lim_{\lambda\rightarrow\infty}&&
{\cal P}({\bf x}(\lambda^2t_1)\in\lambda {\cal D}_1,
{\bf x}(\lambda^2t_2)\in\lambda {\cal D}_2,\ldots,
{\bf x}(\lambda^2t_n)\in\lambda {\cal D}_n|{\bf x}(0)=0)\nonumber\\
&&=\int_{{\cal D}_1}d{\bf x}_1\int_{{\cal D}_2}d{\bf x}_2\ldots
\int_{{\cal D}_n}d{\bf x}_n
p({\bf x}_1,t_1)p({\bf x}_2-{\bf x}_1,t_2-t_1)\ldots
p({\bf x}_n-{\bf x}_{n-1},t_n-t_{n-1})\label{e:multi}
\end{eqnarray}
where $0<t_1<t_2<\ldots<t_n$ are arbitrary times and the ${\cal D}_k$
($k=1,2\ldots n$)
are arbitrary subsets of the plane.  $p({\bf x},t)$ is the Gaussian Green's
function of the (generally anisotropic) diffusion equation in two dimensions,
\begin{equation}\label{e:Gauss}
p({\bf x},t)=\frac{e^{-u^2/4D_{uu}t-v^2/4D_{vv}t}}{4\pi\sqrt{D_{uu}D_{vv}}t}
\end{equation}
and determines
the probability density of a displacement $(\lambda u,\lambda v)$ in time
$\lambda^2t$ in the limit $\lambda\rightarrow\infty$.  The scaling of the
position and time by $\lambda$ is based on the fact that for Gaussian
diffusion the mean square displacement of the particle is proportional to
the time, and is a macroscopic limit.  Here, $u$ and $v$ are distances
measured in the eigendirections of the diffusion tensor, while $D_{uu}$
and $D_{vv}$ are the corresponding eigenvalues.
In the isotropic case $D_{uu}=D_{vv}=D$ and the choice of orthogonal
directions $(u,v)$ is arbitrary.

The components of the diffusion tensor in an arbitrary orthonormal basis
$\{{\bf e}_j\}$ (so that ${\bf x}=\sum_jx_j{\bf e}_j$) can be related to the
mean square displacement $(i,j=1,2)$:
\begin{equation}\label{e:msd}
D_{ij}=\lim_{t\rightarrow\infty}\frac{\langle \Delta x_i\Delta x_j
\rangle}{2t}
\end{equation}
where $\Delta {\bf x}={\bf x}(t)-{\bf x}(0)$.  This expression 
involves the usual 2-time distribution function, that is, $n=1$ in
Eq.~(\ref{e:multi}). The mean square displacement is discussed 
in Ref.~\cite{DC}, where
we found numerically that both chaotic and nonchaotic models exhibited
Gaussian diffusion.  It is conceivable that a system may have Gaussian
2-time distribution functions but non-Gaussian multi-time distribution
functions, i.e., for larger values of $n$.  The results of this note show
that also for larger $n$, both chaotic and nonchaotic models appear diffusive
within the accuracy of the numerical methods.
 
\section{Models}
The models we consider are the same as some of those discussed in
detail in Ref.~\cite{DC}.%
\footnote{Specifically: the periodic circles were denoted LP4
in~\protect\cite{DC}, the randomly positioned circles L$\infty$, the periodic
squares RP4 and the randomly positioned squares R$\infty$.  Here, the L stands
for the Lorentz gas, R for the random orientations 
of the squares, P for periodic, 4 for the linear dimension of the unit cell
(there are also 4 scatterers per unit cell),
and $\infty$ for the infinite ``unit cell''.}
As mentioned above, they all contain a point
particle colliding with fixed non-overlapping scatterers in two dimensions.
The density is such that the scatterers cover exactly half of the
available area.

The microscopically chaotic models are Lorentz models, consisting of
circular scatterers distributed in position either periodically 
or randomly over the plane,
such that they do not overlap.  They are microscopically chaotic as
they have a positive KS entropy.  See Fig.~\ref{f:d}

The microscopically nonchaotic models are Ehrenfest-like models, consisting
of square scatterers distributed in position either periodically 
or randomly over the plane, such that they do not overlap.
They are microscopically nonchaotic, with zero KS entropy.  Unlike the
original Ehrenfest model, the squares in this paper are always oriented 
at random. See Fig.~\ref{f:d}.

\begin{center}
\begin{minipage}{6in}
\begin{figure}
\begin{picture}(400,200)
\put(-80,-530){\epsfig{file=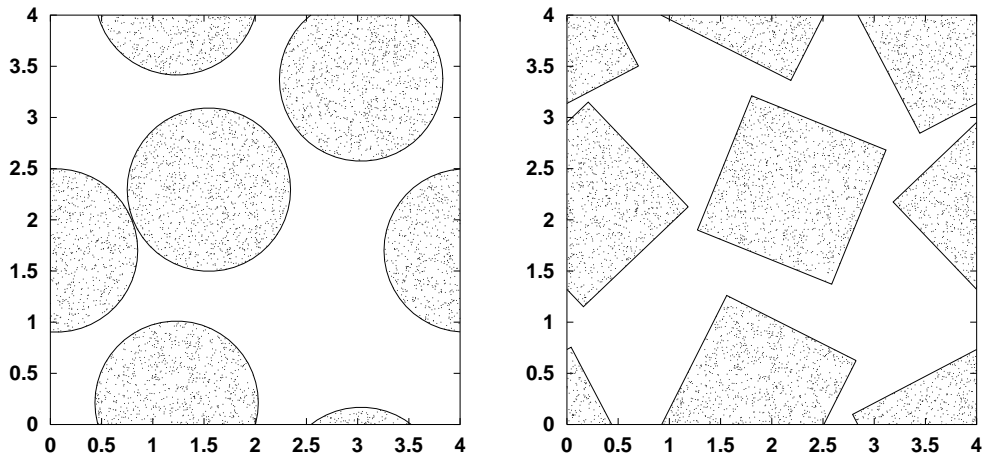, width=600pt}}
\end{picture}
\caption{The periodic Lorentz model (left) and periodic modified Ehrenfest
model (right).
\label{f:d}}
\end{figure}
\end{minipage}
\end{center}

In both the Lorentz and Ehrenfest periodic models, the elementary cell
is square and contains four scatterers.  The random models have no preferred
directions, so the diffusion is isotropic.  In contrast, the periodic
models  
have a directionality defined by the elementary cell, and the diffusion is
anisotropic as in Eq.~(\ref{e:Gauss}).

We investigated Eq.~(\ref{e:multi}) for equally spaced times and
two choices of the subsets ${\cal D}_k$ of the plane:

\begin{description}
\item[A] The ${\cal D}_k$ ($k\leq n$) are all balls of radius
$\epsilon=1/\sqrt{10}$ centered on the origin for $n=1\ldots 7$.
\item[B] $n=2$, where ${\cal D}_1^{(i)}$ and ${\cal D}_2^{(i)}$
are two balls of radius $\epsilon=1/\sqrt{10}$ centered on randomly chosen
points ${\bf x}_k^{(i)}$; $i=1\ldots7$ identifies the different members
of our sample of seven configurations of the ${\cal D}_k^{(i)}$,
see Table~\ref{t:conf}.
\end{description}

\begin{center}
\begin{minipage}{6in}
\begin{table}
\begin{tabular}{crrrrc}
$i$&$x_1^{(i)}$&$y_1^{(i)}$&$x_2^{(i)}$&$y_2^{(i)}$&
Fig.~\protect\ref{f:b} symbol\\
\hline
1& 0.41867&-1.48764&-0.02481&-1.57956&plus\\
2& 1.59402& 0.87882& 1.60281& 1.68137&cross\\
3&-1.08755& 1.54166&-1.54202& 0.53732&star\\
4& 0.90683&-0.60837& 1.05771& 1.39554&open square\\
5&-0.13815&-0.56351& 2.23801&-0.43826&filled square\\
6& 2.19919& 1.04709& 1.55045& 0.49136&open circle\\
7&-0.85062&-0.70816&-2.74100&-2.03641&filled circle
\end{tabular}
\caption{Positions defining the centers of the ${\cal D}_k^{(i)}$ for
case {\bf B}, and corresponding symbols in Fig.~\protect\ref{f:b} below.
\label{t:conf}}
\end{table}
\end{minipage}
\end{center}

In both cases the scale factor varied as $\lambda=1\ldots100$, expanding
the times by $\lambda^2$ and the subsets ${\cal D}_k$ by $\lambda$ as
required by Eq.~(\ref{e:multi}).  In case {\bf A} this means that
as $\lambda$ increases
the ${\cal D}_k$ become larger circles centered on the origin, while in case
{\bf B} the ${\cal D}_k$ become larger but also move off to infinity.
The time steps were a single time unit, so that $t_k=k$ ($k=1\ldots n$),
and the maximum length
of a trajectory was $\lambda^2t_n=70000$ for $\lambda=100$ and $n=7$.
We took 1000 trajectories of length $10^6$ with different initial conditions.
In order to get reasonable statistics without excessive computer time we
considered of order $10^6$ segments of length $7\times10^4$ along each
trajectory, shifting the initial state on the trajectory by one time
unit.  The sample size was thus of order $10^9$.
Because these trajectory segments overlap and hence are not independent,
the statistical error is difficult to estimate a priori; we find below
that it is greatest for the periodic square model, which is not surprising
given that it is expected to have the largest amount of correlation.

In {\bf A} we are testing for the probability for the particle to return
$n$ times at regular intervals to near its starting point.  The motivation
is that while in nonchaotic systems, a trajectory can follow an almost
periodic trajectory for a long time, leading to a power law decay of
recurrences, in chaotic systems the decay of recurrences is exponential.
The power law for nonchaotic systems could lead to correlations over
long times, inconsistent with the uncorrelated diffusion of
Eq.~(\ref{e:multi}).  The integrand in Eq.(\ref{e:multi}) is a product,
so the successive time steps are conditionally independent.
\footnote{Case {\bf A} is similar to the ``almost periodic recurrence''
method of Ref.~\protect\cite{DC}, which showed a difference between
chaotic and nonchaotic models by looking at the probability for return
to near the starting point many times at regular intervals.  The difference
is that in Ref.~\protect\cite{DC} the size of the regions remained fixed
as the time of return increased, while in this paper the ${\cal D}_k$ and
the times are scaled by factors $\lambda$ and $\lambda^2$, respectively.}

In {\bf B} we consider alternative choices for the ${\cal D}_k$ in the
(we think) unlikely event that the rather special case {\bf A} has different
long time behavior than the general case.  Case {\bf B} does not extract
almost periodic orbits but it may reveal other ways in which correlations
might be manifest at long times.
\footnote{It would be somewhat impractical to go beyond $n=2$ in case {\bf
B}:
(a) the integral on the RHS would be of increasingly 
high dimension and hence hard to evaluate, 
and (b) the probabilities would be much lower
so that the statistics would be very poor, see Fig.~\ref{f:b} below.}

\section{Results}
We measured the following quantities for the cases {\bf A} and {\bf B}
discussed above:
\begin{enumerate}
\item The probability ${\cal P}(n;\lambda;\{{\cal D}_k\})$ for various
$\lambda$, $n$ and ${\cal D}_k$ as discussed above.  We often omit some of
the
arguments for simplicity.
\item For case {\bf A}, the conditional probability
${\cal P}_c(n+1|n;\lambda;\{{\cal D}_k\})={\cal P}(n+1)/{\cal P}(n)$ for
the particle being in $\lambda{\cal D}_{n+1}$ given that it was in all
the previous $\lambda{\cal D}_k$ for $k=1\ldots n$.  Again, we often omit
arguments for simplicity.
\end{enumerate}

Eq.~(\ref{e:multi}) actually makes two testable assertions: Does the limit
on the left hand side (LHS) exist, and if so, is it equal to the right
hand side (RHS)?
The above measured quantities give information about the LHS and we are
first interested in the large $\lambda$ limit of these quantities;
for the RHS we obtain the diffusion tensor from Eq.~(\ref{e:msd})
(see Table~\ref{t:diff})
and perform the integration numerically.  This has only been carried out
up to
$n=2$ (which is a four-fold integration).  However, for larger $n$ and
case {\bf A}, the right hand side is (to a very good approximation)
multiplied
by a constant factor for each additional repetition, 
\footnote{The approximation can perhaps be understood as follows.  If we
perform all but the last ${\bf x}_n$ integration, we are left with an
integral
over a function, say $f_n({\bf x}_n)$.  At the $n+1$ stage, we have $f_{n+1}$
as a linear operator ${\cal L}$ (integral over a Gaussian kernel)
acting on $f_n$.  Assuming ${\cal L}$ is well behaved, the $f_n$ approach a
limiting form $\bar{f}\lambda_0^n$ as $n\rightarrow\infty$ where
$0<\lambda_0<1$ is the leading eigenvalue of ${\cal L}$ and $\bar{f}$ is
the leading eigenfunction.  Performing the last integration,
we find that successive integrals differ by a factor of $\lambda_0$, leading
to a vanishing probability for $n\rightarrow\infty$.} 
so the prediction is that
the conditional probabilities become independent of $n$ for large $\lambda$
and $n$.  In fact our result show this behavior already at the smallest
values of $n$.

\begin{center}
\begin{minipage}{6in}
\begin{table}
\begin{tabular}{crrrrr}
Model&$D_{xx}$&$D_{xy}=D_{yx}$&$D_{yy}$&$D_{uu}$&$D_{vv}$\\
\hline
Random squares&0.1480&-0.0009&0.1480&0.1489&0.1471\\
Periodic squares&0.2265&0.0383&0.1304&0.2400&0.1170\\
Random circles&0.2777&0.0005&0.2763&0.2779&0.2761\\
Periodic circles&0.3087&0.0434&0.2878&0.3429&0.2536
\end{tabular}
\caption{Components of the diffusion tensor computed in original coordinates
$(x,y)$ from Eq.~(\protect\ref{e:msd}),
and then diagonalized to give the $(u,v)$ values.  Note that the randomly
placed and oriented squares as well as the randomly placed circles
are isotropic, with $D_{xy}=0$ and $D_{xx}=D_{yy}$ within errors
of about $0.001$, while the periodic models are anisotropic and satisfy
neither of these conditions.
\label{t:diff}}
\end{table}
\end{minipage}
\end{center}

The results are shown in Figs.~\ref{f:a}-\ref{f:b} and Table~\ref{t:rat}.
For all $n$ and $\{{\cal D}_k\}$ the ${\cal P}(\lambda)$
approach a constant (apart from statistical error) at large $\lambda$,
indicating convergence of the limit in Eq.~(\ref{e:multi}), see
Fig.~\ref{f:a}
for case {\bf A} and Fig.~\ref{f:b} for case {\bf B}.
For case {\bf A}, all the conditional probabilities approach
the same constant indicating conditional independence of the multi-time
probabilities as predicted by diffusion, see Fig.~\ref{f:c};
an equivalent statement is that
on a logarithmic scale, all the curves in Fig.~\ref{f:a} become equidistant
at large $\lambda$.
We have verified numerically that in both cases {\bf A} and {\bf B}
the values of the constants as $\lambda\rightarrow\infty$ are consistent
with their values as given on the RHS of Eq.~(\ref{e:multi}), thus
confirming the diffusive nature of all these models.

\begin{center}
\begin{minipage}{6in}
\begin{table}
\begin{tabular}{crrrrr}
Model&${\cal P}_c(2|1)_{th}$&${\cal P}_c(2|1)$&${\cal P}_c(3|2)$&${\cal
P}_c(4|3)$&${\cal P}_c(5|4)$\\
\hline
Random squares&0.1443&0.1457&0.1448&0.1437&0.1426\\
Periodic squares&0.1288&0.1392&0.1295&0.1113&0.1147\\
Random circles&0.0827&0.0831&0.0835&0.0810&0.0825\\
Periodic circles&0.0781&0.0785&0.0788&0.0780&0.0776
\end{tabular}
\caption{Conditional probabilities ${\cal P}_c(n+1|n)$.  The second,
``theoretical'' column gives the values calculated from numerical integration
of the RHS of Eq.~(\protect\ref{e:multi}).
The other columns, derived from Fig.~\protect\ref{f:c} by averaging over
$\lambda$ from 50 to 100, agree with the values from column 2,
and do not depend on $n$, consistent with conditional independence in
the underlying probability distribution.  The differences in the
periodic square model are almost certainly due to statistical errors, as in
Figs.~\protect\ref{f:a},~\protect\ref{f:c}.
\label{t:rat}}
\end{table}
\end{minipage}
\end{center}

\begin{center}
\begin{minipage}{6in}
\begin{figure}
\begin{picture}(400,500)
\put(-50,-300){\epsfig{file=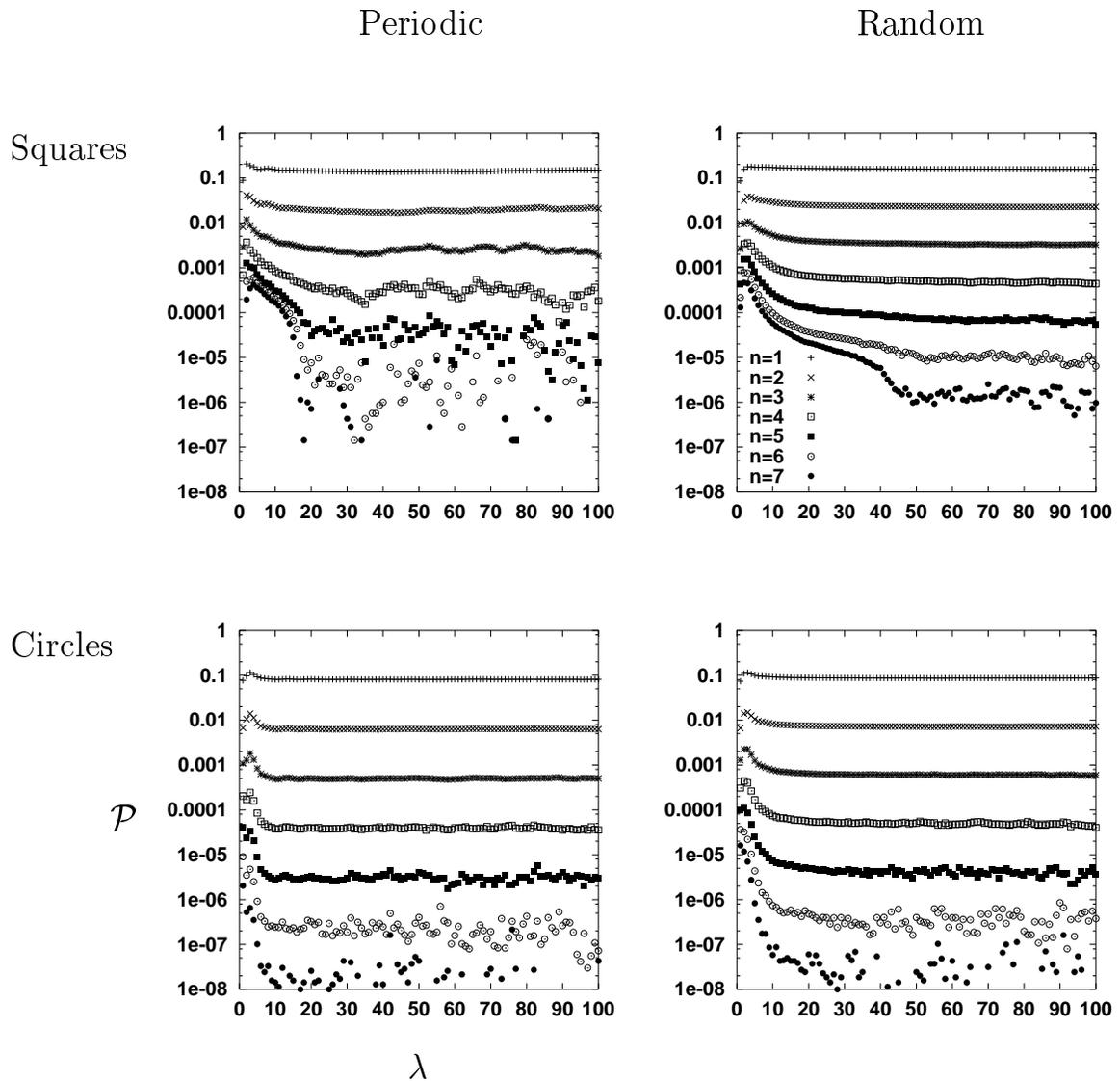, width=600pt}}
\end{picture}
\caption{Probabilities ${\cal P}(\lambda)$ as a function of $\lambda$
for case {\bf A}, with different curves giving $n=1\ldots 7$.
Note that the probability becomes independent
of $\lambda$ indicating the existence of the limit on the LHS of
Eq.~(\protect\ref{e:multi}). The equal spacing between subsequent values
of $n$ indicates a fixed ratio of successive recurrences
since the scale is logarithmic.  This is
consistent with the uncorrelated diffusion as given by the RHS of
Eq.~(\protect\ref{e:multi}).  The statistical error is greatest when the
probability is low, and when the correlations in the data are greatest,
that is, for the periodic square model.\label{f:a}} 
\end{figure}
\end{minipage}
\end{center}

\begin{center}
\begin{minipage}{6in}
\begin{figure}
\begin{picture}(400,500)
\put(-50,-300){\epsfig{file=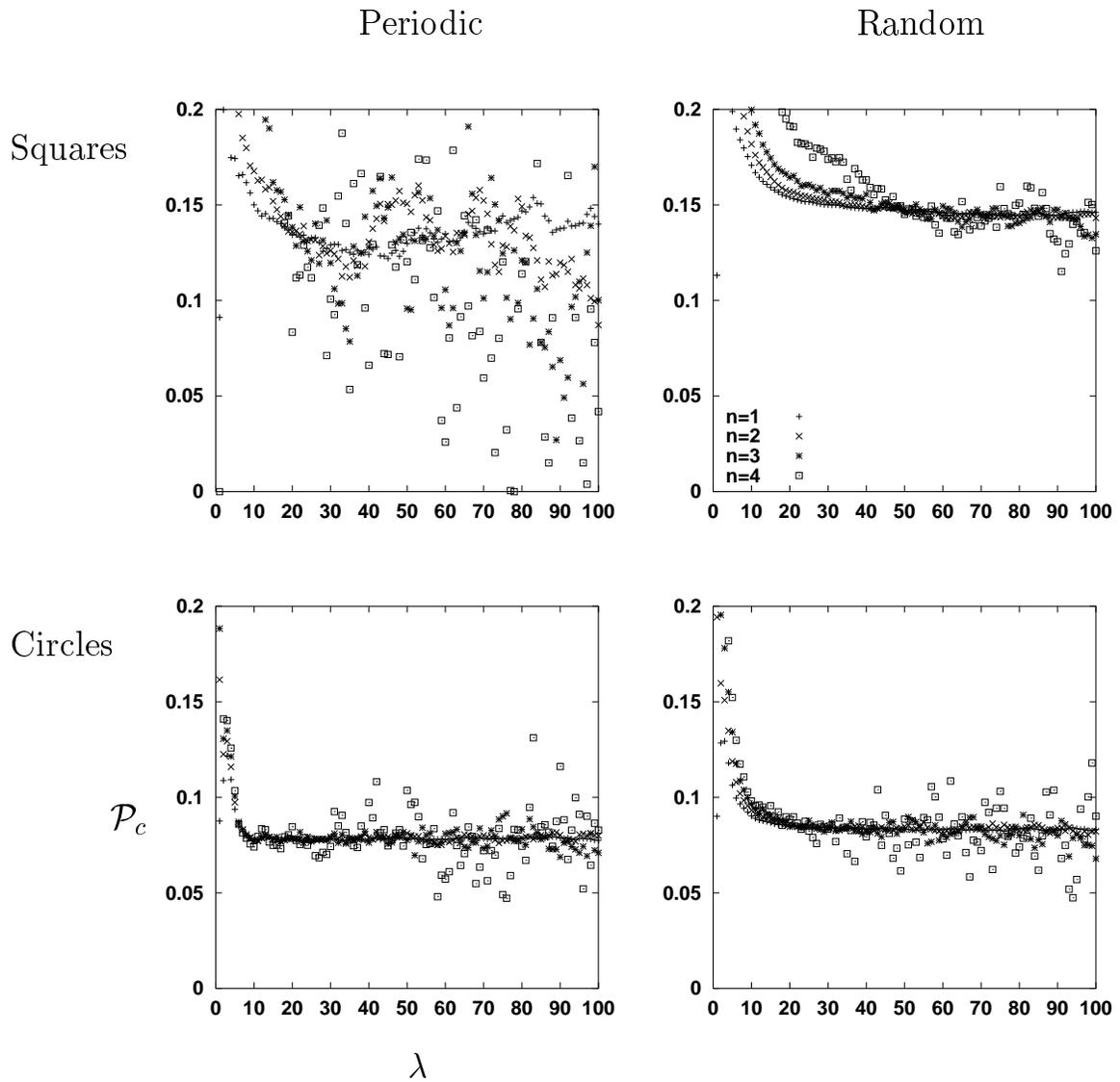, width=600pt}}
\end{picture}
\caption{Conditional probabilities ${\cal P}_c(n+1|n;\lambda)$ as a function
of $\lambda$ for case {\bf A}, and $n=1\ldots 4$, derived from
Fig.~\ref{f:a}.
The equal spacing in Fig.~\protect\ref{f:a} now
causes the graphs corresponding to different values of $n$ to coincide.
Again, statistical error is greatest with low probability (larger $n$)
and with greater correlations in the data (the periodic square model).
\label{f:c}}
\end{figure}
\end{minipage}
\end{center}

\begin{center}
\begin{minipage}{6in}
\begin{figure}
\begin{picture}(400,500)
\put(-50,-300){\epsfig{file=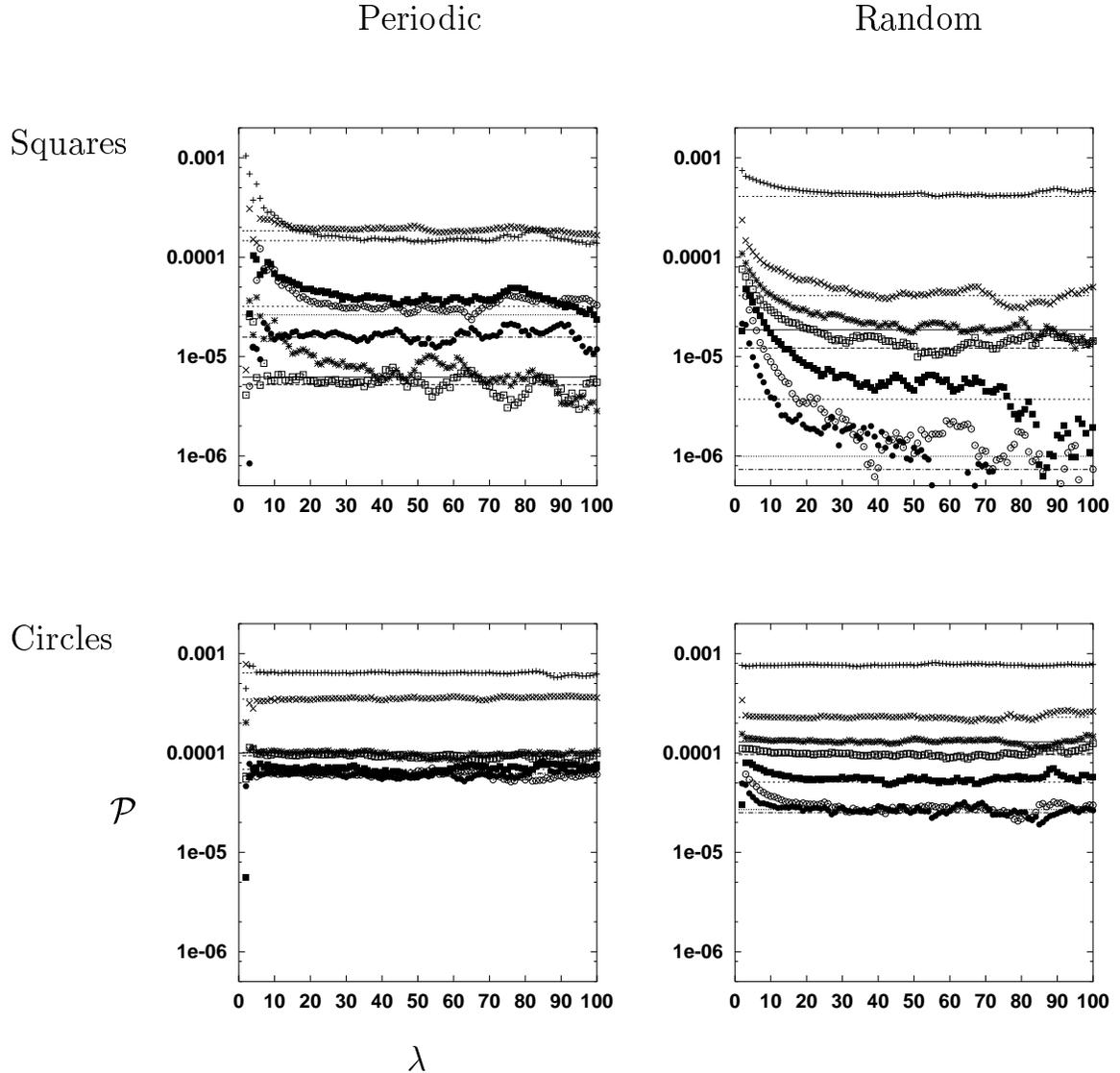, width=600pt}}
\end{picture}
\caption{Probabilities ${\cal P}(\lambda)$ as a function of $\lambda$
for case {\bf B}.  The different symbols correspond
to different ${\cal D}_k^{(i)}$ ($i=1\ldots7$)
as given in Table~\protect\ref{t:conf}.  The
dashed or dotted horizontal lines give the predicted values from the RHS of
Eq.~(\protect\ref{e:multi}), and are not visible when they run through
the data.  The order of the curves differ due to the anisotropy of the
diffusion in the periodic models.
\label{f:b}}
\end{figure}
\end{minipage}
\end{center}

\section{Discussion}
\begin{enumerate}
\item The above mentioned results suggest that there is no difference in the
diffusive behavior of chaotic (Lorentz) and non-chaotic (Ehrenfest) models.
Both exhibit Gaussian diffusion and their only difference is the value of
the diffusion coefficients.
\item The case {\bf A}, in particular the larger probabilities for
recurrence at small $\lambda$ as seen in Fig.~\ref{f:a}, especially
suggests the presence of ``small'' periodic
orbits, which favor return to previously visited regions for the nonchaotic
wind-tree models since they exhibit only algebraic $1/t$ decay
(see Ref.~\cite{DC}).  For the 
chaotic Lorentz models, periodic orbits disappear exponentially fast in $t$
and therefore more returns are less likely.  Nevertheless this difference in
periodic orbit stability, which indicates a more subtle chaotic difference
than the corresponding KS entropies in these models, do not appear to
lead to a qualitatively different, non-Gaussian, diffusion process.
\item The influence of small almost periodic orbits is clearly local and
incoherent from region to region.  It is therefore constantly wiped out
and does not influence the global diffusion process at all, except for
example in the values of the $D_{ij}$, which should decrease by a
strong influence of almost periodic orbits.
\item Our results here appear to differ from the behavior observed in
open (that is, finite with absorbing boundaries) Lorentz and Ehrenfest models
in Ref.~\cite{DC}.  There, we studied the number of particles with different
initial conditions remaining in the system as a function of time.  Only
the Lorentz model exhibited the exponential decay of particles characteristic
of the diffusion equation; the Ehrenfest model corresponding to our squares
exhibited $1/t$ decay due to trapping of particles around periodic orbits.
This is not in contradiction to our results here (for which the nonchaotic
model also exhibits diffusion) since the limits are different: With absorbing
boundaries, the limit $t\rightarrow\infty$ occurs with the linear dimension
of the observed region (the whole system) $L$ fixed;
in this note the linear dimension of the observed
regions $\lambda{\cal D}$ is scaled as $t^{1/2}$, which is
more natural for diffusion.  Thus, the identification of a model with diffusion
depends not only on the microscopic dynamics, but also on the manner in which
the limit is taken.  It would be interesting to investigate these
relationships further.
\item To conclude, our results indicate that at least for the models we
consider, subtle correlations, whether associated with almost periodic
orbits in small clusters of scatterers or otherwise, are insufficient to
modify the long time Gaussian form of the diffusion process when the
natural macroscopic limit is taken (such that lengths are scaled with
the squareroot of times).
In so far as these correlations are determined by the microscopic chaotic
nature of the system, the chaoticity does not seem to show up in the
macroscopic diffusive behavior.  This behavior is consistent with the results
of the Brownian motion experiment discussed in Ref.~\cite{Gaspard}, and the
argument~\cite{DCvB} that the long measuring times of $1/60 {\rm s}$
means that the results could come equally from
chaotic or nonchaotic microscopic dynamics, and that any experimental
determination of microscopic chaoticity would necessarily involve
measurements at microscopic time scales.
\section*{Acknowledgments}
The authors are grateful to a referee of a previous paper~\cite{DC} for
making
the important suggestion discussed in this paper.  They also acknowledge
the support of the Engineering Research Program of the Office of Basic Energy
Sciences of the US Department of Energy under contract number
DE-FG02-88-ER13847.
\end{enumerate}


\begin{references}
\bibitem{Gaspard}P. Gaspard, M. E. Briggs, M. K. Francis, J. V. Sengers,
R. W. Gammon, J. R. Dorfman, and R. V. Calabrese, Nature {\bf 394}, 865
(1998).
\bibitem{DCvB}C. P. Dettmann, E. G. D. Cohen and H. van Beijeren, Nature
{\bf 401}, 875 (1999).
\bibitem{DC}C. P. Dettmann and E. G. D. Cohen, J. Stat. Phys. (in press);
nlin.CD/0001062
\bibitem{LPS}J. L. Lebowitz and J. K. Percus, Phys. Rev. {\bf 155}, 122 (1967);
J. L. Lebowitz, J. K. Percus and J. Sykes, Phys. Rev. {\bf 171}, 224 (1968).
\bibitem{R}R. J. Rubin, J. Math. Phys. {\bf 1}, 309 (1960);
ibid {\bf 2}, 373 (1961).
\bibitem{GLA}S. Goldstein, J. L. Lebowitz, and M. Aizenman, Lecture Notes
in Phys. {\bf 38}, 112 (Springer, Berlin, 1975).
\bibitem{referee}J. R. Norris, ``Markov chains'' (Cambridge University Press,
1997).
\end{references}
\end{document}